\documentclass[twocolumn,showpacs,aps,prb,preprintnumbers,amsmath,amssymb,floatfix]{revtex4}

\usepackage{graphicx}
\usepackage{dcolumn}
\usepackage{bm}
\usepackage{epsfig}

\begin{document}
\date{\today}

\title{ Anomalous galvanomagnetism, cyclotron resonance and microwave spectroscopy of topological insulators }
\author{ G. Tkachov and E. M. Hankiewicz}
\affiliation{ Institute for Theoretical Physics and Astrophysics, University of W\"urzburg, Germany}


\begin{abstract}
The surface quantum Hall state, magneto-electric phenomena and their connection to axion electrodynamics 
have been studied intensively for topological insulators. 
One of the obstacles for observing such effects comes from nonzero conductivity of the bulk. 
To overcome this obstacle we propose to use an external magnetic field to suppress the conductivity of the bulk carriers. 
The magnetic field dependence of galvanomagnetic and electromagnetic responses of the whole system shows anomalies 
due to broken time-reversal symmetry of the surface quantum Hall state, which can be used for its detection. 
In particular, we find {\em negative linear} dc magnetoresistivity and a {\em quadratic} field dependence of the Hall angle, 
{\em shifted} rf cyclotron resonance, {\em nonanalytic} microwave transmission coefficient and {\em saturation} of 
the Faraday rotation angle with increasing magnetic field or wave frequency.
\end{abstract}
\maketitle

\section{Introduction}
\label{Intro}

Unlike ordinary band insulators, semiconductors or semimetals, a recently identified class of materials - 
topological insulators (TIs)~\cite{Kane05,Bernevig06,Koenig07,Fu07,Moore07,Hasan10} - exhibit unusual conducting states on sample boundaries.
On the surface of a three-dimensional TI such a state is characterized by a nodal spectrum with a single Dirac cone (or, in general, 
with odd number of Dirac cones). If time-reversal symmetry (TRS) is broken, an energy gap $\Delta$ is induced at the Dirac points,
and the surface state exhibits the anomalous quantum Hall (QH) effect.~\cite{Qi08,Essin09,Tse10a,Tse10b}
Although the generation of a sizable Dirac gap requires an effort, the surface quantum Hall state in TIs is of great interest because 
it gives rise to rich magneto-electric phenomena~\cite{Qi08,Essin09,Tse10a,Tse10b,Maciejko10,Garate10}
specific to axion electrodynamics.~\cite{Wilczek87}

There is however a serious obstacle for identifying the surface-related magneto-electric phenomena in three-dimensional TIs. 
It stems from dissipative bulk conductivity which generally cannot be ignored because of the complex band structure of three-dimensional TI 
where the Fermi level does not necessarily lie in the bulk band gap or crosses both the surface and bulk states.~\cite{Chen09,Kulbachinskii99,Ayala10,Ren10}
For a TI film with thickness $d$, bulk zero-field dc conductivity $\Sigma$ and surface QH conductivity
$\sigma_{_H}\sim e^2/h$, the contribution of the surface with respect to the bulk is characterized by parameter $e^2/hd\Sigma$.~\cite{Tse10b} 
It has been shown that the well-resolved surface magneto-electric effects, such as the Kerr or Faraday rotation, 
require sufficiently large values of $e^2/hd\Sigma$.~\cite{Qi08,Maciejko10,Tse10b}
If, however, the bulk conductivity is much larger than the surface one, i.e. $e^2/hd\Sigma \ll 1$, 
is it still possible to resolve surface magneto-electric effects in TIs? In this paper we demonstrate such a possibility 
on several different examples of electrodynamic phenomena.     

We show that the surface contribution to the electrodynamics of TIs becomes more pronounced when
the bulk conductivity is suppressed by an external magnetic field ${\bf B}$ and by finite frequency $\omega$ of an applied ac electromagnetic field. 
This can be seen from Boltzmann transport theory expressions 
for the longitudinal $\Sigma_{_L}$ and transverse (Hall) $\Sigma_{_H}$ bulk conductivities (see e.g. Refs.~\onlinecite{LAK56,Fawcett64}):
\begin{eqnarray}
\Sigma_{_L} = \Sigma \frac{1 - i\omega\tau}{  [1 - i\omega\tau]^2  + \Omega^2\tau^2},
\Sigma_{_H} =\Sigma \frac{\Omega\tau}{[1 - i\omega\tau]^2  + \Omega^2\tau^2},
\label{S_LH}
\end{eqnarray}
where $\omega$ and the cyclotron frequency $\Omega=eB/mc$ are both assumed much smaller than the frequency
$\Delta/\hbar$ associated with the surface Dirac gap $\Delta$:
\begin{eqnarray}
\omega, \Omega \ll \Delta/\hbar,
\label{low}
\end{eqnarray}
$m$ is the effective cyclotron mass and $\tau$ is the elastic scattering time. Clearly, with increasing $|\Omega|\tau$ and $\omega\tau$, the real parts of conductivities $\Sigma_{_{L,H}}$ can be made comparable with $e^2/hd$, even though for the zero-field dc case $e^2/hd\Sigma < 1$. Under these conditions the TRS breaking on the TI surface leads to anomalous galvanomagnetic and electromagnetic responses of the {\em whole} system.
In particular, we find (i) negative linear dc magnetoresistivity and Hall angle quadratic with $B$,
(ii) rf cyclotron resonance at shifted frequency
\begin{eqnarray}
\omega_{res} =d\Sigma/\tau|\sigma_{_H}| + |\Omega|,
\label{Om_res}
\end{eqnarray}
(iii) nonanalytic $B$-dependence of the microwave transmission coefficient and (iv)
saturation of the Faraday rotation angle with increasing magnetic field or wave frequency.
Below we explain in detail how these anomalies are related to the surface QH state and how they can be used for its experimental identification.

The paper is organized as follows. In Sec. \ref{Problem} we formulate the main equations of electrodynamics of a TI film and 
discuss approximations used throughout the paper. Then we present the solutions of this electrodynamic problem for different physical situations: 
galvanomagnetic phenomena (Sec. \ref{Galvano}), cyclotron resonance (Sec. \ref{Res}), and electromagnetic transmission and Faraday rotation effects 
(Sec. \ref{Micro}). Finally, in Sec. \ref{Sum} we summarize. 

\section{Formulation of the problem}
\label{Problem}

\begin{figure}[t]
\begin{center}
\includegraphics[width=44mm]{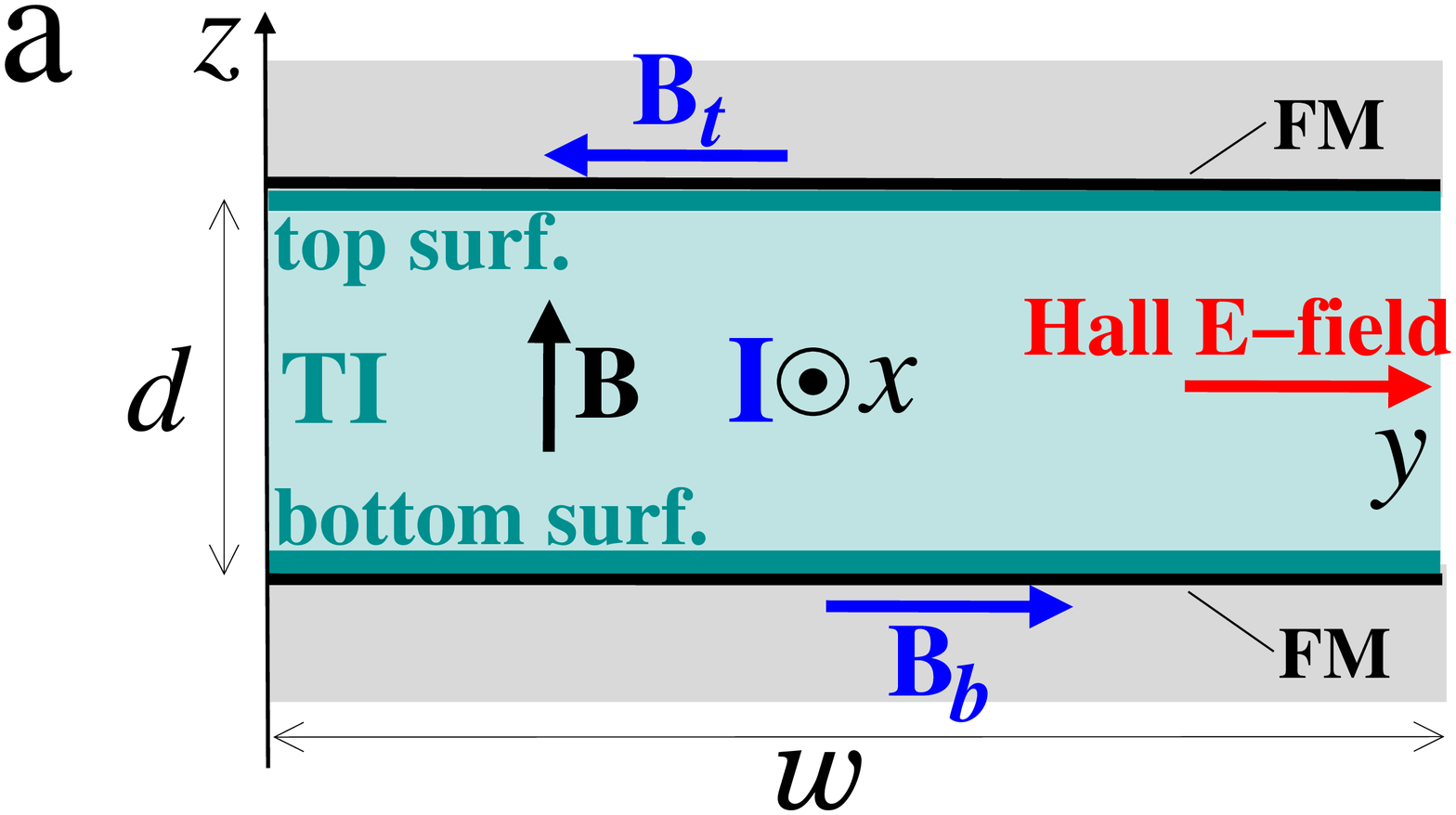}
\hskip 0.5cm
\includegraphics[width=53mm]{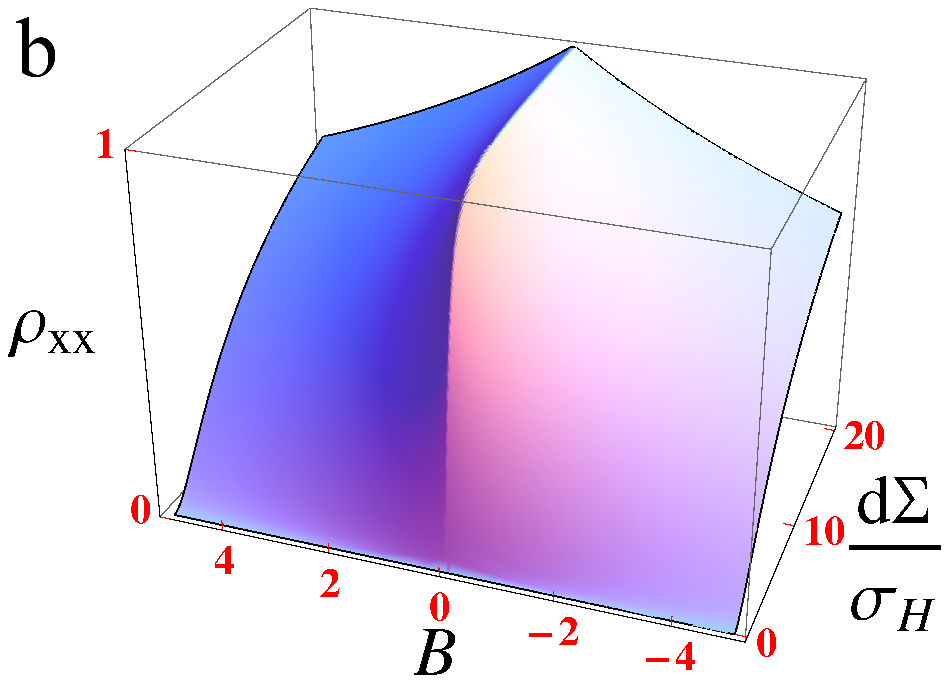}
\includegraphics[width=50mm]{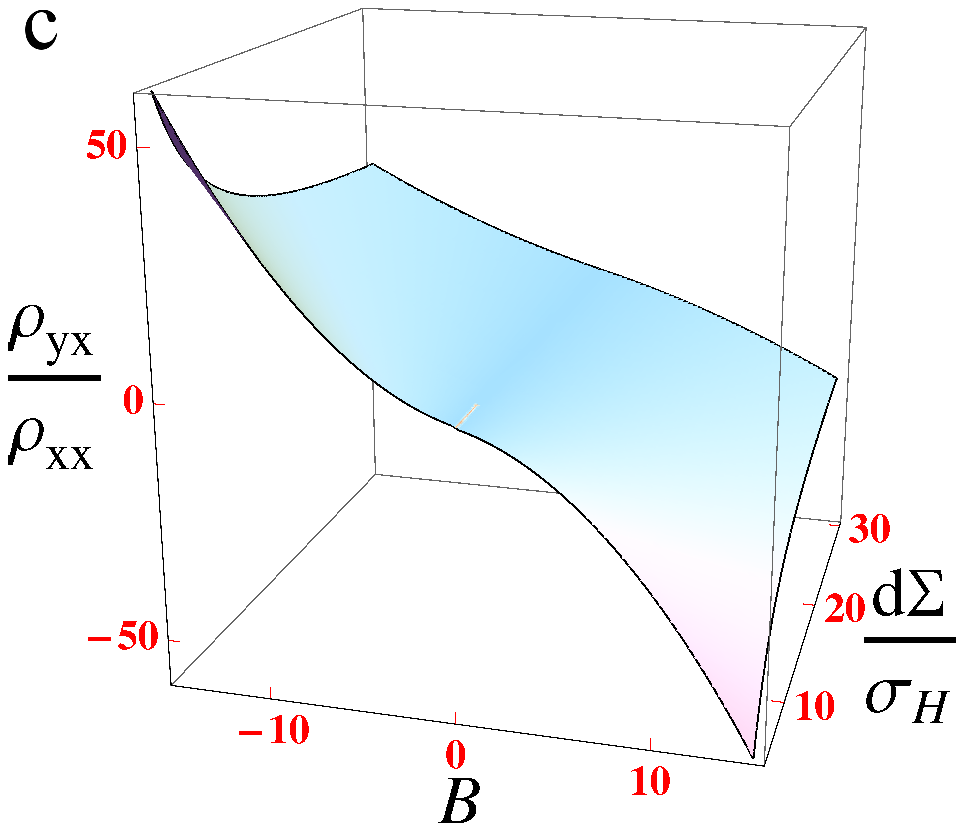}
\end{center}
\caption{
(a) Schematic geometry of a galvanomagnetic experiment with a TI film subject to perpendicular dc magnetic field
${\bf B}$ and electric current ${\bf I}$. 
Magnetic fields ${\bf B}_{t,b}$ at the outer top and bottom surfaces are generated by the current (see, text).
dc longitudinal resistivity $\rho_{xx}$ (b) and Hall angle $\rho_{yx}/\rho_{xx}$ 
(c) versus external magnetic field $B$ and normalized bulk conductivity $d\Sigma/\sigma_{_H}$;
$\rho_{xx}$ and $B$ are in units of $1/\Sigma$ and $mc/e\tau$, respectively. Gray regions are dielectric media.
}
\label{Current}
\end{figure}

TRS breaking on the surface of a TI can be achieved by coating it with thin layers of a ferromagnetic (FM) material,~\cite{Qi08,He10,Zhu10} 
magnetized perpendicularly to the TI film plane by an external dc magnetic field ${\bf B}$ (see, Fig.~\ref{Current}a). 
The FM magnetization acts on the electron spin, generating an energy gap $\Delta$ at the Dirac point, 
which can be described by a Hamiltonian $H=(-1)^s v\mbox{\boldmath$\sigma$}{\bf p} + \Delta\sigma_z$,~\cite{Tse10a} 
where $\sigma=(\sigma_x, \sigma_y, \sigma_z)$ are the spin Pauli matrices, and $v$ and ${\bf p}$ are the velocity and momentum for top $s=+1$ (t) 
and bottom $s=-1$ (b) surfaces. It is assumed that the Fermi level lies within the gap $\Delta$ 
such that both surfaces of the TI have vanishing dissipative longitudinal conductivities and nonzero quantized Hall conductivities:~\cite{Streda82} 
$\sigma^{t,b}_{_H}=e^2\nu_{t,b}/h$, with half-integer filling factors $\nu_{t,b}$.~\cite{Qi08,Tse10b}
If the variation of $B$ is restricted by Eq.~(\ref{low}), the surface states remain on the QH plateaus, 
and $\sigma^{t,b}_{_H}$ do not depend on $|B|$.~\cite{Strong}

We also assume that the surface states respond to a time-dependent electromagnetic (EM) field ($\propto {\rm e}^{-i\omega t}$)  
{\em adiabatically}. This is justified for low frequencies $\omega\ll\Delta/\hbar$ which can, at the same time, be much smaller than the plasma frequency (see below). 
In particular, the surface states remain dissipationless,~\cite{Qi08,Tse10a} i.e.   
the surface current density induced by the electric field ${\bf E}({\bf r})$ can be written as 
\begin{eqnarray}
{\bf j}_{_S}({\bf r}) = [\sigma^t_{_H}  \delta(z-d/2) + \sigma^b_{_H}  \delta(z+d/2)] {\bf\hat z}\, \times {\bf E}({\bf r}),
\label{j_S}
\end{eqnarray}
where ${\bf\hat z}$ is the unit vector perpendicular to the film. The use of the delta functions in Eq.~(\ref{j_S}) 
is justified if the penetration lengths of the surface states into the bulk is much smaller than the film thickness $d$. 
Also, in this case there is no magnetically induced energy gap in the interior of the film. 
Therefore, the bulk conductivity tensor contains both the dissipative (longitudinal) $\Sigma_{_L}$ and Hall $\Sigma_{_H}$ 
components which are both $\omega$-dependent and given for a single (lightest) carrier group 
by the Boltzmann transport theory expressions~(\ref{S_LH}). The resulting bulk current density is
\begin{eqnarray}
{\bf j}_{_B}({\bf r}) = \Sigma_{_L} {\bf E}({\bf r})  + \Sigma_{_H} \, {\bf\hat z} \times {\bf E}({\bf r}).
\label{j_B}
\end{eqnarray}
To find the EM field inside the TI, we use the thin-film approximation $d \ll c/\omega$, 
which for the upper frequency limit $\hbar\omega=\Delta=10$ meV implies $d \ll c\hbar/\Delta\approx 20$ $\mu$m.
In addition, the film thickness, $d$, should be smaller than the skin penetration depth, $\delta$: 

\begin{equation}
d \ll \delta={\rm Re}\sqrt{ \frac{ic^2}{4\pi\omega}\frac{1-i\omega\tau}{\Sigma}  }=\frac{c}{\omega_p} {\rm Re}\sqrt{1+\frac{i}{\omega\tau}},   
\label{skin}
\end{equation}
where $\omega_p=4\pi ne^2\upsilon_{_F}/p_{_F}$ is the plasma frequency of the bulk carriers with density $n$, Fermi velocity $\upsilon_{_F}$ and momentum
$p_{_F}$. For $n=10^{14}$ cm$^{-3}$ and $\upsilon_{_F}=0.5 \cdot 10^6$ ms$^{-1}$ the plasma frequency is $\omega_p\approx 10^{13}$ s$^{-1}$, 
yielding the lower bound for the skin depth $\sim c/\omega_p=0.3$ $\mu$m [see Eq. (\ref{skin})]. Note also that the plasma frequency $\omega_p\approx 10^{13}$ s$^{-1}$  
is of the order of the frequency $\Delta/\hbar\approx 1.5\cdot 10^{13}$ s$^{-1}$ related to the surface gap $\Delta=10$ meV. 
Therefore, in addition to requirement  $\omega\ll\Delta/\hbar$ we have $\omega\ll\omega_p$.

Under condition (\ref{skin}), the electric field ${\bf E}$ inside the film can be approximated by
the average value $\langle {\bf E} \rangle =\int^{d/2}_{-d/2} {\bf E}\, dz/d$.
The equation for $\langle {\bf E} \rangle$ is obtained by averaging the Maxwell equation
$
\nabla\times {\bf B}({\bf r}) + (i\epsilon\omega/c){\bf E}({\bf r}) =
(4\pi/c)[{\bf j}_{_B}({\bf r})  + {\bf j}_{_S}({\bf r})]
$
over the film thickness ($\epsilon$ is the dielectric constant):
\begin{eqnarray}
&&
d\left( \Sigma_{_L} - i\epsilon\omega/4\pi \right)\langle {\bf E} \rangle +
 \left(d\Sigma_{_H} + \sigma_{_H}           \right) {\bf\hat z} \times \langle {\bf E} \rangle =
\nonumber\\
&&
=(c/4\pi)\, {\bf \hat z} \times ( {\bf B}_t - {\bf B}_b), \,\,\,
\sigma_{_H}=(e^2/h)\,(\nu_t +\nu_b).
\label{Eq_E}
\end{eqnarray}
The external EM perturbation enters via magnetic fields ${\bf B}_{t,b}$ at the outer top and bottom surfaces of the TI. 
We specify ${\bf B}_{t,b}$ in each concrete situation considered below.

\section{Galvanomagnetic phenomena}
\label{Galvano}

We begin by considering dc galvanomagnetic phenomena in the standard four-contact geometry
(see, e.g. Ref.~\onlinecite{Kulbachinskii99} and Fig.~\ref{Current}a) in the presence of perpendicular
magnetic field ${\bf B}$ and electric current ${\bf I}$. The current induces the jump of the magnetic field across the film, 
${\bf \hat z} \times ( {\bf B}_t - {\bf B}_b)=(4\pi/cw) {\bf I}$,
so that Eq.~(\ref{Eq_E}) determines the longitudinal and Hall electric fields in terms of given $I$:
$
\langle E_x \rangle = \rho_{xx} \, I/dw
$
and
$
\langle E_y \rangle =\rho_{yx} \, I/dw,
$
where the longitudinal $\rho_{xx}=\Sigma_{_L}/(  \Sigma^2_{_L} + [\Sigma_{_H} + \sigma_{_H}/d]^2 )$
and Hall $\rho_{yx}=-\rho_{xx} [\Sigma_{_H} + \sigma_{_H}/d]/\Sigma_{_L}$ resistivities are given by
\begin{eqnarray}
&&
\rho_{xx} =\frac{1}{\Sigma}
\left[ \left(1 + \frac{\sigma_{_H}}{d\Sigma}\Omega\tau \right)^2 + \left( \frac{\sigma_{_H}}{d\Sigma} \right)^2
\right]^{-1},
\label{rho_xx}\\
&&
\frac{\langle E_y \rangle}{\langle E_x \rangle}=\frac{\rho_{yx}}{ \rho_{xx} }=-\Omega\tau - \frac{ \sigma_{_H} }{d\Sigma} (\Omega^2\tau^2 + 1).
\label{Hall_angle}
\end{eqnarray}
For $\sigma_{_H}=0$ we recover the usual results of the magnetotransport theory:  $\rho_{xx} =1/\Sigma$ and
$\langle E_y \rangle/\langle E_x \rangle=-\Omega\tau$. 
The details of the method used in our calculations are very well described in literature (see, e.g. Refs.~\onlinecite{LAK56,Fawcett64}). 
The $B$-field independent resistivity $\rho_{xx} =1/\Sigma$ reflects strong cyclotron drift in the direction of the current ($x$),
induced by the crossed magnetic $B$ and electric Hall $E_y$ fields.
Moreover, this conclusion remains valid in the nonlinear electrodynamics where the dependence of the bulk conductivity on the magnetic field of 
the current (or of an external EM wave) is taken into account.~\cite{Makarov98}

However, according to Eq.~(\ref{Hall_angle}) the nonzero surface conductivity generates an additional Hall field $\propto \sigma_{_H}/d\Sigma$  
that affects the cyclotron drift in the direction of the current. For this reason
$\rho_{xx}$ (\ref{rho_xx}) acquires the magnetic field dependence.
Moreover, since on the QH plateau $\sigma_{_H}$ does not change with $|B|$,
resistivity $\rho_{xx}(B)$ exhibits linear behavior, whereas the Hall angle
(\ref{Hall_angle}) has the anomalous quadratic $B$ term [Figs.~\ref{Current}b and c].
We note that the anomalous terms in $\rho_{xx}$ and $\rho_{yx}$ remain identifiable even if 
several bulk carrier groups are taken into account, because in that case the bulk resistivities 
are still regular analytic functions of $B$ which can be subtracted from the total $\rho_{xx}$ and $\rho_{yx}$.

\begin{figure}[t!]
\includegraphics[width=70mm]{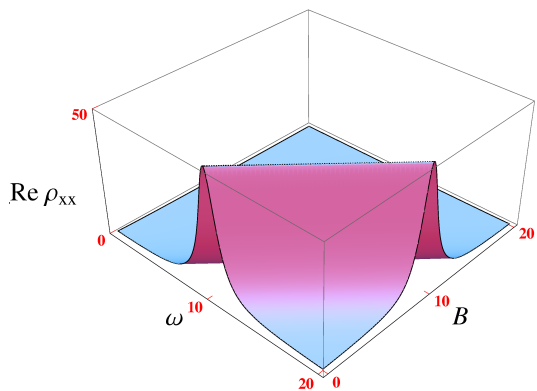}
\caption{
Cyclotron resonance:
Real part of longitudinal ac resistivity $\rho_{xx}$ [see, Eq.~(\ref{rho_xx_om})] versus frequency $\omega$ and magnetic field $B$ (in units of $1/\Sigma$, $1/\tau$ and  $mc/e\tau$, respectively);
$d\Sigma/\sigma_{_H}=10$.
}
\label{CR}
\end{figure}

\section{Cyclotron resonance}
\label{Res}

Let us consider now the bulk cyclotron resonance.
It can be realized in a contactless setup where the sample is placed in the maximum of the electric field of an rf resonator normal mode, 
which generates an antisymmetric magnetic field across the sample, i.e. ${\bf B}_t=-{\bf B}_b$.
At the rf frequencies the displacement current contribution $\epsilon\omega/4\pi$ is usually smaller than
conductivity $\Sigma_{_L}$, so that ${\bf B}_t$ induces an average ac current density
$\langle {\bf j} \rangle=c \, {\bf \hat z} \times {\bf B}_t/2\pi d$ in a contactless way.
The relevant observable is the longitudinal ac resistivity which we find from Eq.~(\ref{Eq_E}) as
\begin{eqnarray}
\rho_{xx}(\omega) =
\frac{ (1-i\omega\tau)/\Sigma }{
\left( 1 + \frac{ \sigma_{_H} }{d\Sigma}\, \Omega\tau \right)^2 + \left(\frac{\sigma_{_H} }{d\Sigma}\right)^2(1-i\omega\tau)^2
}.
\label{rho_xx_om}
\end{eqnarray}
For $\omega\tau > 1$ it has a resonance in both $\omega$ and $B$ dependencies, shown in Fig.~\ref{CR}. Note that the linear on-resonance relation between $\omega$ and $B$ is the hallmark of the cyclotron resonance.
However, the resonant frequency (\ref{Om_res}) is shifted with respect to $|\Omega|$ because of the additional drift in the direction of the current, 
induced by the surface contribution to the Hall electric field $\propto \sigma_{_H}/d\Sigma$ in Eq.~(\ref{Hall_angle}). 
For bulk Drude conductivity $\Sigma=ne^2\upsilon_{_F}\tau/p_{_F}$
the frequency shift depends on the bulk carrier density $n$, Fermi velocity $\upsilon_{_F}$ and momentum
$p_{_F}=\hbar k_{_F}$ as
\begin{eqnarray}
\omega_{res} =2\pi n d \upsilon_{_F}/|\nu_t + \nu_b| k_{_F} + |\Omega|.
\label{Om_res1}
\end{eqnarray}
For $n=10^{14}$ cm$^{-3}$, $d=50$ nm, $\upsilon_{_F}=0.5 \cdot 10^6$ ms$^{-1}$ and $|\nu_t + \nu_b|=1$  
the resonance frequency shift is $\omega_{res} - |\Omega|\approx 10^{12}$ s$^{-1}$, 
i.e. well below both $\Delta/\hbar\approx 1.5\cdot 10^{13}$ s$^{-1}$ and $\omega_p\approx 10^{13}$ s$^{-1}$.

\section{Electromagnetic transmission and Faraday rotation}
\label{Micro}

\begin{figure}[t]
\begin{center}
\includegraphics[width=55mm]{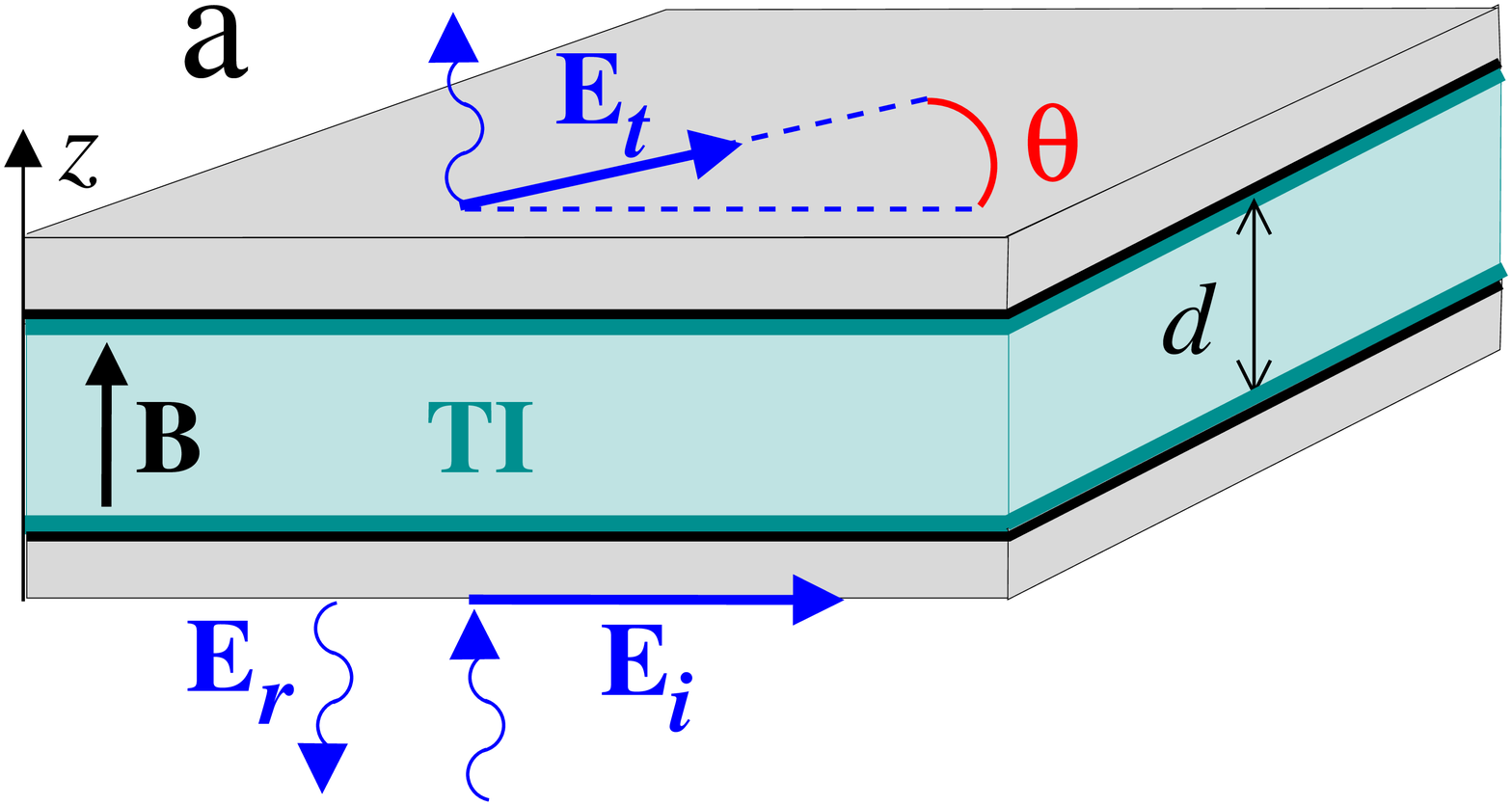}
\vskip 0.2cm
\includegraphics[width=42mm]{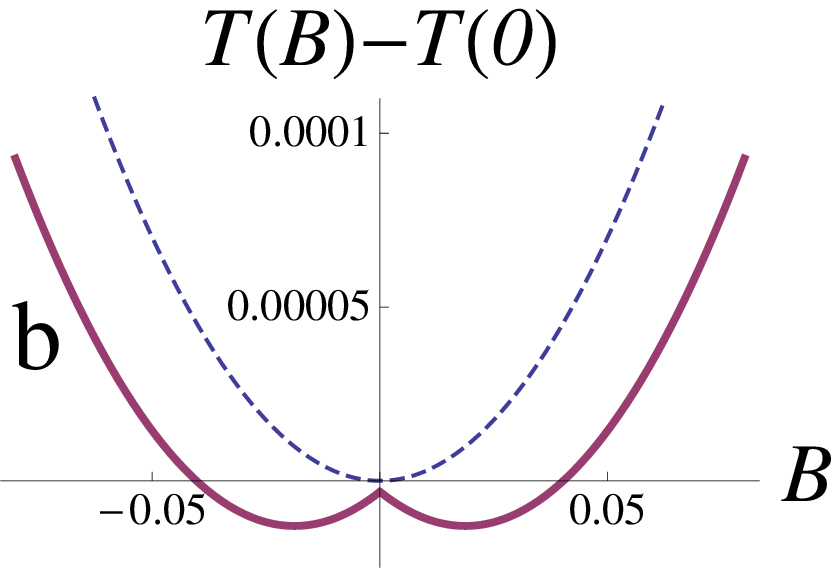}
\includegraphics[width=42mm]{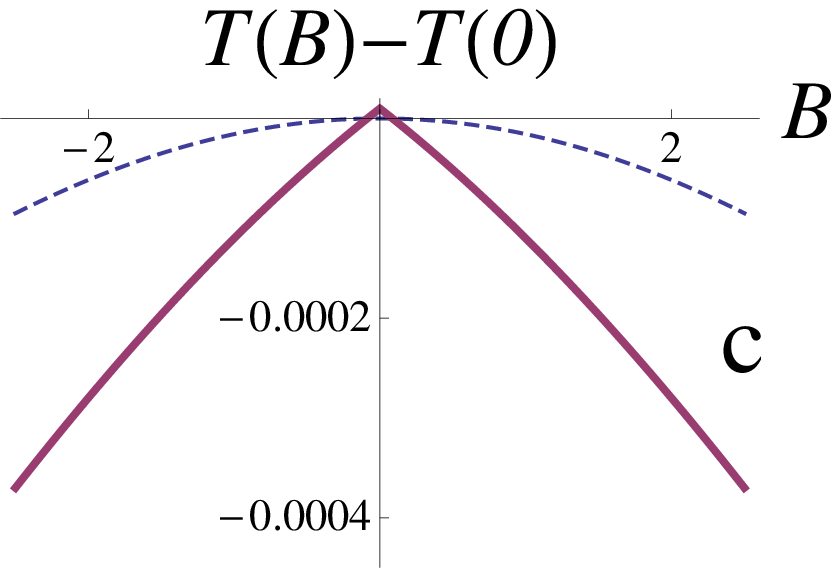}
\vskip 0.2cm
\includegraphics[width=42mm]{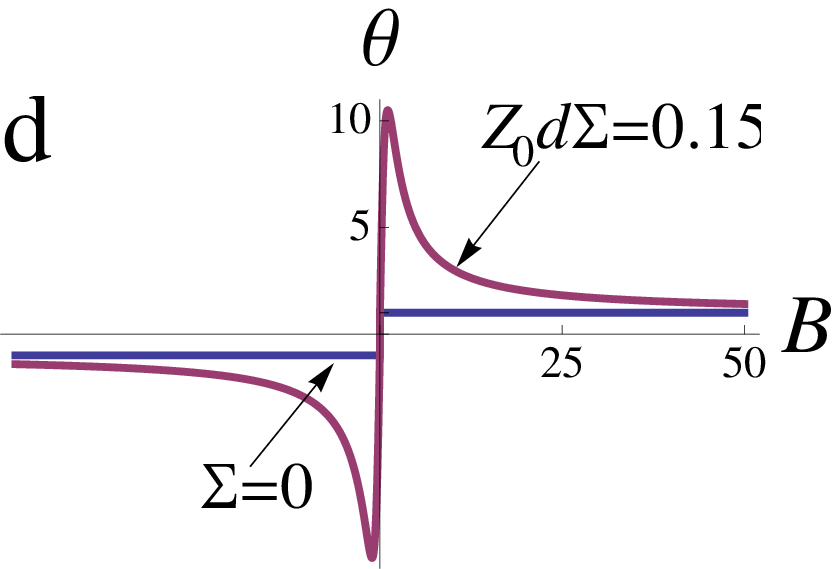}
\includegraphics[width=42mm]{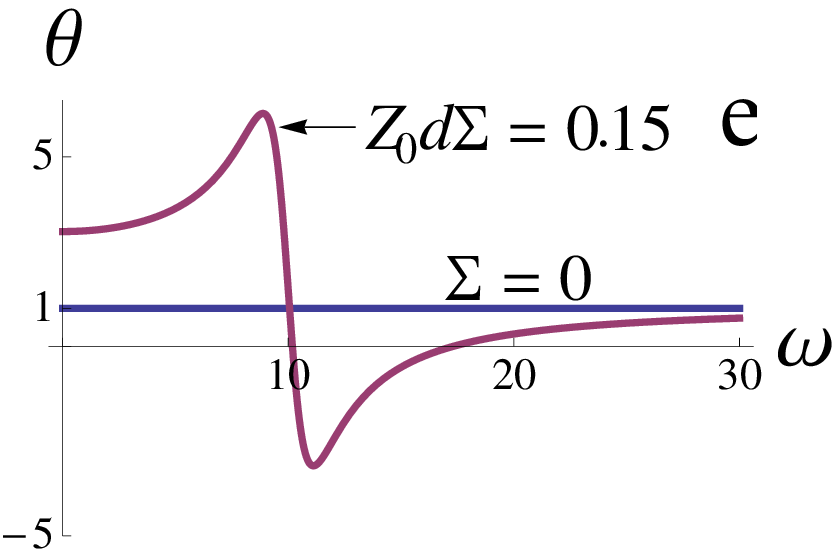}
\end{center}
\caption{
(a) Schematic geometry of EM transmission spectroscopy of a TI film (see also text).
(b) Magnetic-field dependent part of the transmission coefficient for $\sigma_{_H}\not =0$
(solid curves) and for $\sigma_{_H}=0$ (dashed curves);  $\omega\tau=1$ and $Z_0d\Sigma=1$.
(c) Same for $\omega\tau=35$ and $Z_0d\Sigma=40$.
(d) Faraday angle in units of $\alpha=e^2/c\hbar$ versus magnetic field for zero and finite bulk conductivity $\Sigma$; $\omega\tau\ll 1$.
(e) Faraday angle versus frequency for zero and finite bulk conductivity $\Sigma$; $\Omega\tau=10$.
$B$ and $\omega$ are in units of $mc/e\tau$ and $1/\tau$, respectively.
All data for $\epsilon_{t,b}=1$ and thin film with $Z_0d\epsilon\omega/4\pi\sim d\epsilon\omega/c\ll 1$.
}
\label{Faraday}
\end{figure}

We now turn to microwave spectroscopy which also allows one to probe the surface states in TIs~\cite{Ayala10}.
We consider an EM wave incident normally at the bottom surface of a TI and will analyze both the transmission
coefficient and the Faraday rotation of the EM field plane in the transmitted wave [see, also Fig.~\ref{Faraday}a]. 
This situation involves a new conductivity scale, viz. the inverse impedance of the dielectric media surrounding the TI, 
$Z^{-1}_0=c(\sqrt{\varepsilon_t} + \sqrt{\varepsilon_b})/4\pi$, 
where $\varepsilon_{t,b}$ are the dielectric constants of the top and bottom materials (shown in gray in Fig.~\ref{Faraday}a).
The new conductivity scale $Z^{-1}_0$ is important because it is much larger than the surface QH conductivity $\sigma_{_H}$: 
the product $Z_0|\sigma_{_H}|$  is a small parameter proportional to the fine structure constant $\alpha=e^2/c\hbar$~\cite{Qi08,Tse10a,Tse10b}:
\begin{equation}
Z_0|\sigma_{_H}|=2\alpha |\nu_t + \nu_b|/(\sqrt{\varepsilon_t} + \sqrt{\varepsilon_b}) \ll 1.
 \label{Z}
\end{equation}
Therefore, the response of the surface state to an EM wave is generically rather weak.
To proceed we note that the electric and magnetic fields on the outer surfaces of the TI film are
\begin{eqnarray}
&&
{\bf E}_t \approx \langle {\bf E} \rangle, \qquad \qquad \quad
{\bf B}_t =\sqrt{\epsilon_t} \, {\bf\hat z}\times {\bf E}_t,
\label{Fields_CR_t}\\
&&
{\bf E}_b = {\bf E}_i + {\bf E}_r \approx \langle {\bf E} \rangle, \,
{\bf B}_b =\sqrt{\epsilon_b}\, {\bf\hat z}\times ({\bf E}_i - {\bf E}_r),
\label{Fields}
\end{eqnarray}
where $t$ refers to the transmitted wave on the top surface, whereas
$i$ and $r$ label the incident and reflected waves on the bottom ($b$) surface.
In the thin film approximation the electric field on each surface equals to the average field.
Eliminating the reflected field through ${\bf E}_r\approx \langle {\bf E} \rangle - {\bf E}_i$,
we express the magnetic field difference as
$
{\bf B}_t - {\bf B}_b \approx {\bf\hat z} \times
[ (\sqrt{\epsilon_t} + \sqrt{\epsilon_b}) \langle {\bf E} \rangle - 2\sqrt{\epsilon_b}\, {\bf E}_i],
$
insert this into Eq.~(\ref{Eq_E}), and solve it for $\langle {\bf E} \rangle$.
Since the incident electric field ${\bf E}_i$ can be regarded real, we present the solution for the real part of the
transmitted wave:
\begin{eqnarray}
{\rm Re}\,{\bf E}_t \approx {\rm Re}\,\langle {\bf E} \rangle =
T\, ( {\bf E}_i \, \cos\theta + {\bf E}_i \times {\bf \hat z}\, \sin\theta),
\label{E_t}\\
T=\frac{ 2 }{ 1 + \sqrt{\varepsilon_t/\varepsilon_b} }\, \sqrt{a^2_{_L} + a^2_{_H}}, \qquad
\theta=\arctan\frac{ a_{_H} }{ a_{_L} },
\label{T}
\end{eqnarray}
where $T$ is the transmission coefficient, $\theta$ is the rotation angle of the EM field plane with respect to the incident wave (Faraday angle), 
and $a_{_L}(\omega, B)$ and $a_{_H}(\omega,B)$ are real functions given by
\begin{eqnarray}
a_{_L}={\rm Re}
\frac{ 1 + Z_0 d(\Sigma_{_L} - i\epsilon\omega/4\pi) }{ [ 1 + Z_0 d(\Sigma_{_L} - i\epsilon\omega/4\pi) ]^2 + Z^2_0[d\Sigma_{_H} + \sigma_{_H}]^2},\,\,\,
\label{a_L}
\end{eqnarray}
\begin{eqnarray}
a_{_H}={\rm Re}
\frac{ Z_0( d\Sigma_{_H} +\sigma_{_H}) }{ [ 1+ Z_0 d(\Sigma_{_L} - i\epsilon\omega/4\pi) ]^2 + Z^2_0[d\Sigma_{_H} + \sigma_{_H}]^2}.\,\,\,
\label{a_H}
\end{eqnarray}
Similar to resistivity (\ref{rho_xx}), the TRS breaking leads to a nonanalytic linear $B$ dependence of the transmission coefficient (\ref{T}). 
To illustrate this we extract the large zero-field value $T(0)$ from $T(B)$ and plot in Figs.~\ref{Faraday}b and c the difference $T(B)-T(0)$ 
for $\sigma_{_H}\not =0$ (solid curves) and for $\sigma_{_H}=0$ (dashed curves). 
The magnetic field range in which $T(B)-T(0)\propto |B|$ can be tuned by varying parameters $\omega\tau$ and $Z_0d\Sigma$. 
This should help in finding the optimal regime for observation of the predicted anomalous magnetic-field dependence of $T$.

Figure~\ref{Faraday}d shows the low-frequency Faraday angle $\theta$ (\ref{T}) in units of the fine structure constant $\alpha$ 
as a function of the magnetic field $B$ for zero and finite bulk conductivity $\Sigma$.
For $\Sigma=0$ the Faraday angle contains only the surface contribution
$\theta\approx Z_0\sigma_{_H}=2\alpha (\nu_t +\nu_b)/( \sqrt{\varepsilon_t} + \sqrt{\varepsilon_b}) \propto {\rm sgn}(B)$~\cite{Qi08,Tse10a,Tse10b}. 
For $\Sigma\not =0$ the bulk contribution makes the dependence $\theta(B)$ nonmonotonic with the following asymptotics:
\begin{eqnarray}
\theta \approx \frac{ Z_0\sigma_{_H} }{  1 + Z_0d\Sigma },\,\, \Omega\tau\ll 1, \qquad
\theta \approx Z_0\sigma_{_H},\,\, \Omega\tau\gg 1,
\label{asympts}
\end{eqnarray}
The low-field limit ($\Omega\tau\ll 1$)  agrees with the result of Ref.~\onlinecite{Tse10b} which found $\theta$ smaller 
than the surface contribution $Z_0\sigma_{_H}$ for nonzero bulk conductivity $\Sigma$, e.g.
$\theta\approx \sigma_{_H}/d\Sigma \ll Z_0\sigma_{_H}$ for $Z_0d\Sigma\gg 1$.
However, for strong fields $\Omega\tau\gg 1$ we find saturation of $\theta(B)$ precisely at the surface value $Z_0\sigma_{_H}$ 
because of the suppression of the bulk conductivity via classical cyclotron motion.
This way of extracting the topological surface contribution may have an advantage over the previously proposed low-field detection scheme~\cite{Qi08} 
because the magnetization of the FMs in strong fields leads to a more robust surface Dirac gap $\Delta$. As seen from Fig.~\ref{Faraday}e, 
the frequency dependence of the Faraday angle $\theta(\omega)$ also saturates at the surface value $Z_0\sigma_{_H}$, 
which can be used for its detection as well.

\section{Summary}
\label{Sum}

In summary, we have investigated galvanomagnetic and electromagnetic properties of topological insulators in which
time-reversal symmetry is broken due to the surface quantum Hall effect. 
Our model includes both the dissipationless quantum Hall conductivity on the surface and the classical magnetoconductivity in the bulk of the system.
Although the zero-field dc bulk conductivity may significantly exceed the surface one, 
the surface contribution can still be detected through anomalous magnetic field dependencies of electrodynamic responses, 
revealing the underlying broken time-reversal symmetry. With appropriate modifications our findings can be extended to 
HgTe quantum wells which also support single-valley Dirac fermions~\cite{Buettner11,GT11a,GT11b,GT11c,GT10} and show a pronounced Faraday effect.~\cite{Shuvaev11}  

\acknowledgments
We thank A. H. MacDonald, S.-C. Zhang, L.W. Molenkamp, A. Pimenov, A. M. Shuvaev and G. V. Astakhov for helpful discussions. 
The work was supported by DFG grant HA5893/1-1.


\begin{thebibliography}{99}
\bibitem{Kane05}
C. L. Kane and E. J. Mele, Phys. Rev. Lett. {\bf 95}, 226801 (2005).

\bibitem{Bernevig06}
B. A. Bernevig and T. L. Hughes and S. C. Zhang, Science {\bf 314}, 1757 (2006).

\bibitem{Koenig07}
M. K\"onig, S. Wiedmann, C. Br\"une, A. Roth, H. Buhmann, L. W. Molenkamp, X.-L. Qi, and S.-C. Zhang, Science {\bf 318}, 766 (2007).

\bibitem{Fu07}
L. Fu, C. L. Kane and E. J. Mele, Phys. Rev. Lett. {\bf 98}, 106803 (2007).

\bibitem{Moore07}
J. E. Moore and L. Balents, Phys. Rev. B {\bf 75}, 121306(R) (2007).

\bibitem{Hasan10}
M. Z. Hasan and C. L. Kane, Rev. Mod. Phys. {\bf 82}, 3045 (2010); X.-L. Qi and S.-C. Zhang, e-print arXiv:1008.2026 (unpublished).

\bibitem{Qi08}
X.-L. Qi, T. L. Hughes, and S.-C. Zhang, Phys. Rev. B {\bf 78}, 195424 (2008).

\bibitem{Essin09}
A. M. Essin, J. E. Moore, and D. Vanderbilt, Phys. Rev. Lett. {\bf 102}, 146805 (2009).

\bibitem{Tse10a}
W.-K. Tse and A. H. MacDonald, Phys. Rev. Lett. {\bf 105}, 057401 (2010).

\bibitem{Tse10b}
W.-K. Tse and A. H. MacDonald, Phys. Rev. B {\bf 82}, 161104(R) (2010).

\bibitem{Maciejko10}
J. Maciejko, X.-L. Qi, H. D. Drew, and S.-C. Zhang, Phys. Rev. Lett. {\bf 105}, 166803 (2010).

\bibitem{Garate10}
I. Garate and M. Franz, Phys. Rev. Lett. {\bf 104}, 146802 (2010).

\bibitem{Wilczek87}
F. Wilczek, Phys. Rev. Lett. {\bf 58}, 1799 (1987).

\bibitem{Chen09}
Y. L. Chen, J. G. Analytis, J.-H. Chu, Z. K. Liu, S.-K. Mo, X. L. Qi, H. J. Zhang, D. H. Lu, X. Dai, Z. Fang, S. C. Zhang, I. R. Fisher, Z. Hussain, and  Z.-X. Shen,
Science {\bf 325}, 178 (2009).

\bibitem{Kulbachinskii99}
V.A. Kulbachinskii, N. Miura, H. Nakagawa, H. Arimoto, T. Ikaida, P. Lostak, and C. Drasar,
Phys. Rev. B {\bf 59}, 15733 (1999).

\bibitem{Ayala10}
O. E. Ayala-Valenzuela, J. G. Analytis, J.-H. Chu, M. M. Altarawneh, I. R. Fisher, and  R. D. McDonald,
e-print arXiv:1004.2311 (unpublished).

\bibitem{Ren10}
Z. Ren, A. A. Taskin, S. Sasaki, K. Segawa, and Y. Ando, Phys. Rev. B {\bf 82}, 241306(R) (2010). 

\bibitem{LAK56}
I. M. Lifshitz, M. Ya. Azbel, and M. I. Kaganov, Zh. Eksp. Theor. Fiz. {\bf 31}, 63 (1956) [Sov. Phys. JETP {\bf 4}, 41 (1957)]. 

\bibitem{Fawcett64}
E. Fawcett, Adv. Phys. {\bf 13}, 139 (1964). 


\bibitem{He10}
H.-T. He, G. Wang, T. Zhang, I.-K. Sou, G. K. L. Wong, J.-N. Wang, H.-Z. Lu, S.-Q. Shen, and F.-C. Zhang, Phys. Rev. Lett. {\bf 106}, 166805 (2011). 

\bibitem{Zhu10}
J.-J. Zhu, D.-X. Yao, S.-C. Zhang, and K. Chang, Phys. Rev. Lett. {\bf 106}, 097201 (2011).

\bibitem{Streda82}
P. Streda, J. Phys. C {\bf 15}, L717 (1982).

\bibitem{Strong}
In contrast, Ref.~\onlinecite{Tse10b} considers strong orbital and Zeeman magnetic field effects on the surface and
no magnetic field influence in the bulk.


\bibitem{Makarov98}
N. M. Makarov, G. B. Tkachev, and V. E. Vekslerchik, J. Phys.:  Condens. Matter. {\bf 10}, 1033 (1998).

\bibitem{Buettner11}
B. B\"uttner, C. X. Liu, G. Tkachov, E. G. Novik, C. Br\"une, H. Buhmann, E. M. Hankiewicz, P. Recher, B. Trauzettel, S. C. Zhang and L. W. Molenkamp, 
Nature Phys. {\bf 7}, 418 (2011).

\bibitem{GT11a} 
G. Tkachov, C. Thienel, V. Pinneker, B. B\"uttner, C. Br\"une, H. Buhmann, L. W. Molenkamp, and E. M. Hankiewicz, 
Phys. Rev. Lett. {\bf 106}, 076802 (2011).

\bibitem{GT11b}
G. Tkachov and E. M. Hankiewicz, Phys. Rev. B {\bf 84}, 035444 (2011).

\bibitem{GT11c}
G. Tkachov and E. M. Hankiewicz, Phys. Rev. B {\bf 83}, 155412 (2011). 

\bibitem{GT10}
G. Tkachov and E. M. Hankiewicz, Phys. Rev. Lett. {\bf 104}, 166803 (2010). 


\bibitem{Shuvaev11}
A. M. Shuvaev, G. V. Astakhov, A. Pimenov, C. Br\"une, H. Buhmann, and L. W. Molenkamp,
Phys. Rev. Lett. {\bf 106}, 107404 (2011). 



\end{thebibliography}
\end{document}